\documentclass[%
 aip,
rsi,%
 amsmath,amssymb,
 reprint,%
]{revtex4-1}

\usepackage{xcolor}
\usepackage{graphicx}
\usepackage{dcolumn}
\usepackage{bm}
\usepackage{nicefrac}

\begin{document}

\preprint{AIP/123-QED}

\title[Rodr\'{i}guez et al.]{RF current condensation in the presence of turbulent enhanced transport}

\author{E. Rodr\'{i}guez}
 \altaffiliation[Email: ]{eduardor@princeton.edu}

\author{A. H. Reiman}%
 \altaffiliation[Email: ]{areiman@pppl.gov}
\author{N. J. Fisch}
 \altaffiliation[Email: ]{fisch@princeton.edu}
 \affiliation{ 
Department of Astrophysical Sciences, Princeton University, Princeton, NJ, 08543
}
\affiliation{%
Princeton Plasma Physics Laboratory, Princeton, NJ, 08540
}%

\date{\today}

\begin{abstract}
Sharp temperature gradients in a magnetically confined plasma can lead to turbulent motion of the plasma.
This turbulence in turn enhances the transport of heat across magnetic field lines.  
The enhanced transport impacts the temperature differential that can be sustained in magnetic islands between the island centre and its periphery.
It is shown here that, by limiting this temperature differential, this enhanced transport can have a profound influence on the extent to which the RF current condensation effect stabilises the island growth.
Interestingly, because  the heat transport is no longer simply linear in the temperature gradient, the RF current condensation effect also exhibits entirely new hysteresis phenomena.

\end{abstract}

\maketitle

\section{\label{sec:intro} Introduction:}

A key issue confronted by tokamak reactors is the control of magnetic islands.
These islands, driven unstable by neoclassical tearing modes (NTMs), deteriorate confinement and trigger catastrophic disruptions \cite{Buttery_2000,devries,devries14}.   
Tearing modes were theoretically predicted to be stabilised by RF  currents driven near the island centre \cite{reiman,Yoshioka,hegna}. 
This stabilisation through noninductive RF current drive techniques  has been experimentally demonstrated \cite{Bernabei,Warrick2000,Gantenbein,Zohm2001,Isayama2000,LaHaye2002,Petty04}, 
with continued attention paid to optimising stabilisation \cite{Sauter2004,Kamendje2005,LaHaye2006,Lahaye08,Henderson08,Lazzari,Volpe,Sauter2010,Bertelli2011,Hennen2012,Smolyakov_2013,Ayten,Borgogno,Volpe2015,Fevrier2016,Wang2015,JC_Li_2017,Grasso_JPP2016,Grasso18,Poli15}.

A variety of RF waves can be used to drive noninductive current in tokamaks \cite{fisch87}, but, for stabilising the NTM, the most studied methods are electron cyclotron current drive (ECCD)  \cite{fisch80} and  lower hybrid current drive (LHCD) \cite{fisch78}.  These methods use waves that incur power deposition on the tail of the electron distribution function, drawing out a current of superthermal electrons, and thus leading to high current drive efficiency.   The power deposition is also  extremely sensitive to the electron temperature \cite{karney81,karney79}, which can result at high enough power in the {\it RF current condensation effect} \cite{Reiman18}.  
The condensation of current occurs because of a nonlinear feedback mechanism:  the deposition of energy in a magnetic island raises its temperature, so that, for waves sensitive to the temperature, there is an increased power deposition.  
The temperature increase is largest at the island centre, so that the power deposition and current tend to be maximised also in the island centre, possibly leading to a more stabilising use of the RF current.  

However, if the electron temperature is raised throughout the island, then significant extra wave damping may be incurred near the island periphery.  
The damping near the island periphery leads to increased temperature and damping there, leading to a feedback effect that can effectively \textit{shadow} the island centre, preventing the wave energy from penetrating deep inside the island \cite{rodr19}.  This shadowing effect can be overcome by taking into account the nonlinearity of the power deposition in aiming the ray trajectories, but it can place a premium on accurate aiming of the ray trajectories.

When the temperature enhancement in the island becomes very large, then so will temperature gradients particularly near the periphery.
With these large temperature gradients, the perpendicular heat flux may increase and no longer be simply proportional to the gradient. 
In fact, large temperature gradients are a source of free energy which can trigger various instabilities, which then increase heat transport, leading to what are 
called {\it stiff} temperature profiles \cite{Garbet2004,Maget18,Kotsch95}.  
Thus, it can be expected that the increase in heat transport can limit the temperature gradient near an island periphery, thereby also reducing the temperature. This change might in certain cases set a limitation on current condensation but also reduce the shadowing when relevant (increasing the stabilising effect as RF current drive can condense nearer the centre).

It is the objective of this work to examine these effects resulting from stiffness.  The paper is organised as follows:  
In Sec.~II, we briefly introduce a first form of the model and show that stiffness leads to multiple steady state solutions and a new hysteresis effect, alternative to the hysteresis effect due to the finite available power that was previously explored  \cite{rodr19}.  
In Sec.~III, we consider a more general model for the temperature stiffness, which exhibits a double bifurcation solution, with increased possibilities for hysteresis.  
The exploration of more general models suggests also the possibility of using experimentally observed temperature perturbations to infer the proper stiffness model.
In Sec.~IV, we show how the stiffness affects the wave damping -- and how it might hamper the condensation effect or minimise shadowing.
In Sec.V, we summarise our conclusions.  
Certain additional details are reserved for appendices.

\section{Hysteresis with strong turbulence}

The basic model constructed to explain the effects of current condensation in magnetic islands consists of an energy balance. RF wave power deposition plays the role of the driver, while diffusive losses through the island edge regulate the temperature of the island as it is heated. The model is only concerned with the steady state of such a balance, faster, however, than the resistive growth time of the island. This allows to take the island width, $W_i$, and thus the boundaries of the problem to be fixed. \par
We assume the island width to be small relative to the minor radius of the plasma.  The island is then extremely elongated in the poloidal direction relative to its radial width.  This suggests the use of a 1D slab model of the island interior.  The island centre (O-point) is represented by the surface at x=0, and surfaces at $\pm x$ are taken to correspond to the same flux surface within the island. The heat conduction along the field lines is assumed to be sufficiently fast that the temperature is constant on the flux surfaces within the island, so that the temperature is symmetric about x=0.  This implies that the temperature is constant on the separatrix, which is represented by the surfaces at $x=\pm W_i/2$, where $W_i$ is the island width.\par
This then preserves the basic effects of the change in topology introduced by the island.
A solution of the nonlinear thermal diffusion equation in this simplified model was compared with that in magnetic island geometry in [\onlinecite{Reiman18}] and was found to reproduce the qualitative features of the solution in the more realistic island geometry, as well as to give a rough approximation to the threshold for the nonlinear effects in the more realistic solution. \par
For modelling power deposition, it is paramount to note that the damping in velocity space occurs on the tail of the Maxwellian electron distribution function. The energy transfer from the wave to the plasma is proportional to the size of the resonant superthermal population $\exp[-(v_\parallel/v_{T})^2]\approx\exp(-w^2)\exp(u)$, where $w=v_\parallel/v_{T0}\sim10$ represents a normalised resonant wave velocity, $v_{T0}$ is the background electron thermal velocity,\cite{karney81,fisch87} $u=w^2\widetilde{T}/T_0$ with $\widetilde{T}$ representing the temperature perturbation and $T_0$ the background temperature. For this section in particular, and for simplicity, we will take the power in the wave, unlike the deposition, not to change within the island, as introduced in [\onlinecite{Reiman18}]. For more details on the implicit assumptions and a detailed discussion, we refer the reader to [\onlinecite{Reiman18}] and [\onlinecite{rodr19}].\par
    The energy balance equation may thus be written in the form,
    \begin{equation}
	\frac{d}{d\widetilde{x}}\left(\widetilde{\kappa}\frac{du}{d\widetilde{x}}\right)=-P_0e^u, \label{eqn::modNW}
	\end{equation}
	where $\widetilde{x}=2x/W_i$ is the normalised space coordinate and $\widetilde{\kappa}$ is the thermal conductivity normalised to the non-turbulent background. $P_0$ is a measure of wave power; in fact, the deposition when the non-linear temperature feedback is not considered. Since $u$ measures the temperature differential of the island compared to the separatrix (whose temperature is taken to be fixed), Eqn.~(\ref{eqn::modNW}) is to be solved with homogeneous boundary conditions $\widetilde{x}=\pm1$.\cite{rodr19} \par
	It is assumed that electrons and ions are both characterised by the same temperature, which will describe the case when the temperature equilibration time between electrons and ions is fast compared to the diffusion time.\cite{suying20} 
	\par
	The model, as described, does not include a description of the turbulent enhancement of heat transport, but it might be accommodated through the heat conductivity $\widetilde{\kappa}$. Let $\widetilde{\kappa}$ be generalised from $\widetilde{\kappa}=1$ into a function of temperature gradient. In its simplest functional form, and taking $'$ to denote spatial derivatives
	\begin{equation}
	    \widetilde{\kappa}=\begin{cases}
	          1 & \text{if $u'\leq u'_\mathrm{th}$}\\
              \infty & \text{if $u'>u'_\mathrm{th}$}
    \end{cases} 
	\end{equation}
	Below some temperature gradient threshold, $u'_\mathrm{th}$, heat transport is non-turbulent. This threshold represents a sufficient condition for an instability to occur. The value of the threshold depends on the nature of the instability, and thus the species it belongs to. We make the simplifying assumption $T_e=T_i$, and take a `single fluid' stiffness model, leaving the species dependence for later work. \par
	When an instability takes place beyond the threshold, the model allows turbulent transport to conduct any amount of heat. This divergent $\widetilde{\kappa}$ is of particular mathematical convenience, as it keeps all temperature gradients below $u'_\mathrm{th}$. This is what the solution for a finite $F>u'_\mathrm{th}$ to $\widetilde{\kappa} u'=F(\widetilde{x})$ (from Eqn.~(\ref{eqn::modNW})) yields, where $F(\widetilde{x})=\int_0^{\widetilde{x}}P_0\exp[u(x')]\mathrm{d}x'$. As a result, temperature profiles will have an increased triangular shape. \par
	Taking these results into account, Eqn.~(\ref{eqn::modNW}) may be cast in the form of an eigenvalue equation:
	\begin{subequations}
	\label{eqn::NWeigen}
	\begin{align}
	\lambda=&\cosh^2\left(\sqrt{\lambda P_0/2}\right) ~~~ &\left(\lambda\leq\lambda^*\right)\\
	=&\cosh^2\left(x_\mathrm{E}\sqrt{\lambda P_0/2}\right)\left(\lambda-\lambda_*+1\right) ~ &\left(\lambda>\lambda^*\right)
	\end{align}
	\end{subequations}
	where $x_E=\ln(\lambda -\lambda^*+1)/u'_\mathrm{th}-1$, and $\lambda^*=1+(u'_\mathrm{th})^2/2P_0$ is the eigenvalue corresponding to the non-stiff solution with $\max|u'|=u'_\mathrm{th}$; i.e., the coldest solution for which stiffness is relevant. By construction, $\lambda\leq\lambda^*-1+e^{u'_\mathrm{th}}$; more details may be seen in Appendix A. \par
	\begin{figure}[h]                                                                       
	\hspace*{-0.2cm}
    \includegraphics[width=0.5\textwidth,keepaspectratio]{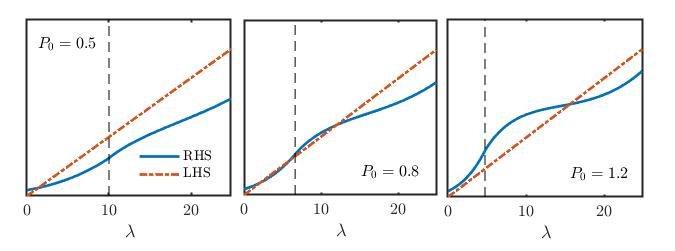}
    \vspace*{-0.5cm}
    \caption{RHS and LHS of the eigenvalue equation (\ref{eqn::NWeigen}) for $u'_\mathrm{th}=3$ and three different powers $P_0$, showing hysteresis. The discontinuity in the derivative occurs at $\lambda=\lambda^*$ (shown as a dashed black line). In this context the threshold for hysteresis at $u'_\mathrm{th}=2$ corresponds to the condition for which the LHS is tangent to the RHS at $\lambda=\lambda^*$. In that case, it is not possible to have three solutions.}\label{fig::fig1}
    \end{figure}
    
	This eigenvalue problem may be analysed using a graphical interpretation (see Fig.~\ref{fig::fig1}). Doing so naturally leads to separating solutions as a function of the gradient threshold into two groups.\par
\begin{figure}[h]
    \includegraphics[width=0.5\textwidth]{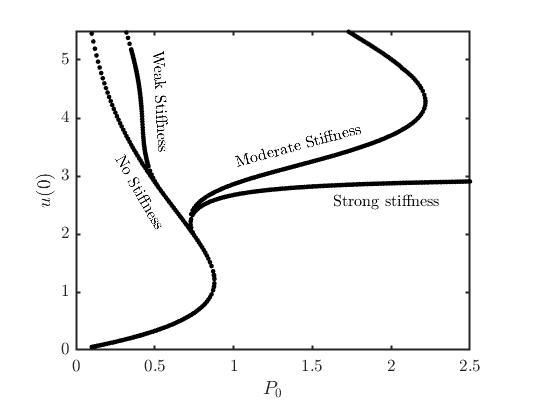}
    \vspace*{-0.7cm}
    \caption{Central island temperature as a function of power for different models, including no stiffness, strong stiffness at $u'_\mathrm{th}=3$, moderate stiffness ($u'_\mathrm{th}=3, \widetilde{\chi}_s=5$) and weak stiffness ($u'_\mathrm{th}=3, \widetilde{\chi}_s=1$). These show the main solution features.}\label{fig::fig2}
    \end{figure}

	For $u'_\mathrm{th}>2$, the solutions for the island temperature show hysteresis behaviour as power is changed. This can be deduced from Fig.~\ref{fig::fig1}, where a graphical representation of the eigenvalue problem for three different powers is shown. Each intersection point represents a temperature profile solution. Thus, from the three points in the central panel, only one at a time could truly describe the island. Which one it will be depends on whether the steady state is reached starting from the left (lower power) or right (larger power) panels. This is the way hysteresis manifests itself, and is perhaps more clearly seen represented as the central temperature curve labelled `strong stiffness' in Fig.~\ref{fig::fig2}. The hysteresis curve is formally very similar to that observed in [\onlinecite{rodr19}], though the temperature saturating mechanism that makes the problem globally stable (i.e. an odd number of solutions for all $P_0$) is different. Let us describe this in some more detail.\par 
	The region with three solutions may be defined as $P_*<P_0<P_b$, where the boundary values $P_*$ and $P_b$ label bifurcations.  The one occurring at $P_b=0.88$ is consistent with that shown in [\onlinecite{Reiman18}]. This bifurcation results from the exponential increase  with temperature of the power deposition overcoming the increase in diffusion losses. The apparent irrelevance of stiffness for $P_b$ is a consequence of all temperature gradients for solutions colder than the bifurcation being below the turbulent threshold. Indeed, the bifurcation point itself corresponds to an edge gradient (the largest across the island) of $u'=2$, consistent with the statement of $u'_\mathrm{th}=2$ being the threshold for the appearence of this $P_b$ bifurcation.  \par
	The upper branch, which exists for $P_0>P_*$, physically represents a balance between the enhancement of edge transport losses and the gains from the additional exponential deposition. Any hotter temperature would lead to an imbalance favouring the loss of energy, placing an obvious upper bound to $u(0)<u'_\mathrm{th}$ (see `strong stiffness' in Fig.~\ref{fig::fig2}). This hotter branch could be accessed starting from a cold island, and increasing $P_0$ gradually to exceed $P_b$. If the power was diminished below $P_b$, the island would remain hot down to $P_*$ due to the enhanced deposition scenario from which we start. 
	\par
    
	When $u'_\mathrm{th}<2$, the saddle-node bifurcation at $P_b$ disappears. It may be shown that, in this case, a single stable solution exists for all values of $P_0$, with no trace of hysteresis. This loss of hysteresis solutions underlines the relevance of the gradient threshold on the structure of the final solution, and ultimately on the possibility of the condensation effect. \par

\section{Condensation with relaxed stiffness: road to double bifurcation}
    The transport model assumed so far has been of mathematical convenience, but of limited physical relevance. It only represents an extreme case; namely, that of limitless heat transport due to turbulence. A more relevant model would instead accommodate a finite level of transport in the turbulent regime. There exist a number of models that do so\cite{Maget18,Kotsch95,Garbet2004}; but, here we will consider the so-called \textit{critical gradient transport} model\cite{Garbet2004}. \par
    The critical gradient model writes the increase in heat conductivity as follows,
    \begin{equation}
	     \widetilde{\kappa}=1+\widetilde{\chi}_s(u'-u'_\mathrm{th})H(u'-u'_\mathrm{th}) \label{eqn::criticalGradTrans}
	\end{equation}
	where $H$ represents the Heaviside function. Though the critical gradient $u'_\mathrm{th}$ has the same meaning as previously, it is usually defined in terms of a dimensionless parameter $\kappa_c$, such that $ u'_\mathrm{th}=\Upsilon\kappa_c$ where $\Upsilon=W_iw^2/2R$ and $R$ the major radius (see Appendix B for more details).\par
	The other new parameter introduced in Eqn. (\ref{eqn::criticalGradTrans}), which constitutes an additional degree of freedom, is $\widetilde{\chi}_s$. This parameter describes the stiffness of the system; i.e., it regulates by how much turbulence enhances transport. Again, it is customary to define it in terms of $\chi_s/\chi_0$ as $\widetilde{\chi}_s=\chi_s/\chi_0\Upsilon$ (see Appendix B). Our previous strong stiffness model corresponds to the limit $\chi_s/\chi_0\rightarrow\infty$, while non turbulent transport (such as [\onlinecite{Reiman18}]) corresponds to $\chi_s/\chi_0\rightarrow0$.\par
    \par
	With this generalised form of $\widetilde{\kappa}$, Eqn.~(\ref{eqn::modNW}) may be solved for numerically. A fundamental distinction between the solution to this model and the previous strong turbulence solution is that the solution space is no longer globally stable. In other words, beyond some critical value of $P_0$ (or for islands hotter than the hottest steady state $u(0)$ solutions), the temperature of the island would grow indefinitely, without converging to any steady state (see `moderate stiffness' curve in Fig.~\ref{fig::fig2}). The lack of such a steady state is, again, the result of the exponential $e^u$ power deposition dominating, this time over the linear increase in the transport losses via $\widetilde{\kappa}$ as the temperature in the island rises. This might be seen formally from the solution to $\widetilde{\kappa}u'=F$, which yields, as $F\rightarrow\infty$, $u'\rightarrow\infty$. The boundless growing temperature is nonphysical\cite{Reiman18}, and it would eventually meet some additional physics or the break-down of assumptions such as the smallness of $\widetilde{T}/T_0$. \par
	Even though the solution space is globally unstable in this model, it is evidently different from the non-stiff case. In fact, the finite stiffness softens the effects of the non-linear feedback (one may compare curves labelled `moderate stiffness' and `no stiffness' in Fig.~\ref{fig::fig2}), increasing the region of stability. \par

	In addition to this fundamental global change, the new degree of freedom $\widetilde{\chi}_s$ is capable of shaping the structure of the solution. In particular, the system may be forced to exhibit either one or two bifurcations (see Fig.~\ref{fig::fig2}). \par
	In the case of a stringent critical temperature gradient threshold, in the same fashion as for the prior very stiff case, the bifurcation at $P_b=0.88$ would be affected. Because stiffness is however weaker, thresholds lower than $u'_\mathrm{th}<2$ are needed for the bifurcation to vanish. Even when the solution might not bifurcate at $P_b$, it will exhibit a new bifurcation at larger power (see Fig.~\ref{fig::fig2}). This is explained by the same mechanism that prevented the solution space from being globally stable, and thus its location depends on the strength $\widetilde{\chi}_s$. \par
	The possibility of a double bifurcation such as the one shown in Fig.~\ref{fig::fig2} is then contingent not only on a large enough gradient threshold so that the $P_b$ bifurcation exists, but also on the magnitude of $\widetilde{\chi}_s$. Weak stiffness (see Fig.~\ref{fig::fig2}) would not introduce such drastic changes but could instead lead merely to a displacement of the solutions into larger $P_0$ by an amount regulated by $\widetilde{\chi}_s$.
	\par
	In summary, this freedom to shape the solutions suggest that some active or passive manipulation of transport properties could be exploited to change the response of condensation. In addition, one could perhaps exploit this intricate form of the solution to experimentally measure stiff properties of the plasma. To do so, one might launch RF waves at different powers and observe the resulting temperature variations in the island, from which, a priori, a particular stiffness model could be fitted.

\section{Interplay between enhanced transport and wave damping in condensation}
    So far, stiffness has been shown to act as an energy loss mechanism which limits more or less effectively the action of nonlinear effects. The limitless accessibility to energy from the wave led however to globally unstable solution spaces for finite stiffness. Thus, we introduce here a more physical finite energy wave damping model, to study its interaction with stiffness and condensation. \par
The wave damping model used is inherited from previous work\cite{rodr19}, to which the reader is referred for further details. Fundamentally, the model describes the RF wave energy density, $V$, as it travels left to right (increasing $\widetilde{x}$) through the island and becomes damped. This damped energy is what drives the temperature diffusion equation (symmetrised to account for the fast flux surface heat conduction). The corresponding ray and diffusion equations are respectively, 
    \begin{subequations}
\begin{align}
V'(\widetilde{x})&=-\alpha_0e^uV(\widetilde{x})  \label{eqn::WaveEnergy} \\
\frac{\mathrm{d}}{\mathrm{d}\widetilde{x}}\left(\widetilde{\kappa}\frac{\mathrm{d}u}{\mathrm{d}\widetilde{x}}\right)&=\frac{V'(\widetilde{x})+V'(-\widetilde{x})}{2} \label{eqn::diffus}
\end{align}
\end{subequations}
    Define the initial value of the wave energy $V(\widetilde{x}=-1)\equiv V_0$ with typical values $V_0\sim1-10$, and $\alpha_0$ to represent the linear damping strength of the wave normalised to the island width, typically $\alpha_0\sim0.1-3$.\cite{rodr19} 
    Note that the definition $\widetilde{x}=2x/W_i$ is kept for these equations unlike in [\onlinecite{rodr19}].
    We again adopt the critical gradient form of $\widetilde{\kappa}$ to introduce the effects of stiffness into the problem.\par
\begin{figure}[h]
   \hspace*{-0.6cm}
    \includegraphics[width=0.5\textwidth]{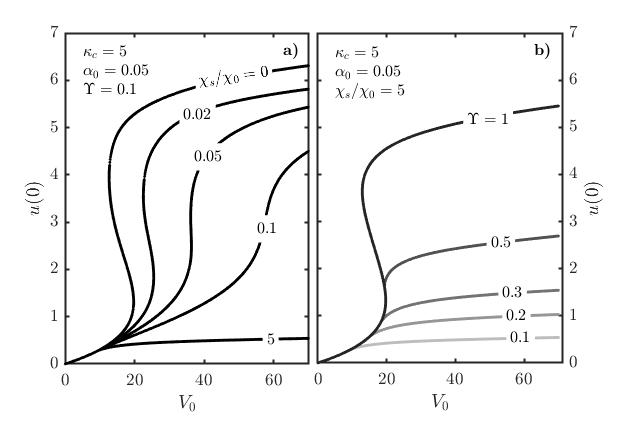}
    \vspace*{-0.7cm}
    \caption{a) Island centre temperature for different stiffness $\chi_s$ values, showing its influence on the resulting solution structure ($\kappa_c=5$, $\alpha_0=0.05$, $\Upsilon=0.1$). Hysteresis disappears at values of $\chi_s/\chi_0\geq 0.05$ b) Island centre temperature for different $\Upsilon$ values, showing the reduced effects of stiffness for larger islands ($\kappa_c=5$, $\alpha_0=0.05$, $\chi_s/\chi_0=5$).  }\label{fig::fig3}
    \end{figure}

    In the line of previous sections, the non-linear effects in the solutions to Eqns.~(\ref{eqn::WaveEnergy}) and (\ref{eqn::diffus}) are found to be moderated by stiffness. Generally, the island becomes colder. The qualitative changes in the  $u(0)-V_0$ solution space and the bifurcations are similar, especially for $\alpha_0\ll1$, to the changes observed in the previous sections (see Figs.~\ref{fig::fig2} and \ref{fig::fig3}a). The exact response of the system is controlled by both the strength of stiffness and the gradient threshold for a given damping strength. Let us dedicate a more detailed look to this.  
    \par 
    The threshold $u'_\mathrm{th}$ roughly delimits the  subspace of solutions in $V_0$ space affected by stiffness. That subset roughly includes all those cases for which the RF power exceeds the value $V_0\gtrsim 2\Upsilon\kappa_c$ (as the cutoff gradient $u'_\mathrm{th}=\Upsilon\kappa_c$, while the non-linear non-stiff edge slope $u'\sim V_0/2$). Within this set, the larger $V_0$, the more significant the effects of stiffness. The characteristic scale is set by the magnitude of $\widetilde{\chi}_s$. \par
    Therefore, the smaller the value of $u'_\mathrm{th}$ and the larger $\widetilde{\chi}_s$ the more noticeable the effects are. Both these parameters scale with $\Upsilon\propto W_i/R$ as seen in Appendix B. This means that smaller islands at larger radii are likely to be more affected (see Fig.~\ref{fig::fig3}b). This suggests that the effects of finite stiffness will be most relevant for situations with smaller islands. If very small, enough for stiffness to dominate the non-linear condensation effects, the latter will not be observable, as may be seen in Fig.~\ref{fig::fig3}.  \par 

    Another important aspect of the solution that should be studied is the self-consistent power deposition profile, in particular for the location of its peak. In fact, power profiles are more directly related to stabilisation of the island than $u(0)$, as we take current drive to be approximately proportional to the deposited power. \par
\begin{figure}[h]
   \hspace*{-.5cm}
    \includegraphics[width=0.55\textwidth,keepaspectratio]{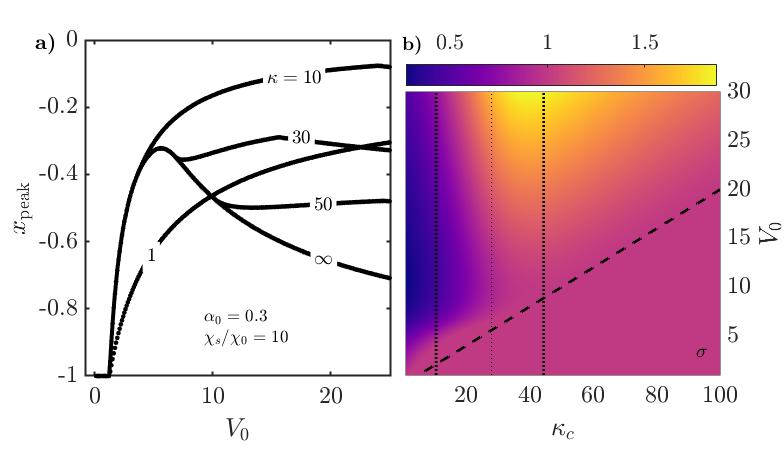}
    \vspace*{-0.7cm}
    \caption{a) Normalised position of the peak of the deposition with input RF power, for a number of stiffness thresholds. These correspond to parameters $\alpha_0=0.3$, $\chi_s/\chi_0=10$ and $\Upsilon=0.1$. b) Changes in stabilisation $\sigma$ as a function of $\kappa_c$ and $V_0$. Normalisation is done with respect to the non-stiff solution for each respective $V_0$ value. The broken line represents the approximate boundary between affected and unaffected solutions by stiffness. The three dotted lines (from left to right) represent roughly the separation of the three groups presented in the main text.
    }\label{fig::fig4}
    \end{figure}

     We introduce an additional measure of RF stabilisation in addition to the location of the deposition peak. Actually, the position of the deposition peak is only a partial account, as its magnitude as well as the amount of energy that leaks the island also play a role. Hence, we introduce parameter $\sigma$ as the figure of merit describing the stabilisation power of a given power deposition profile (see Appendix C). $\sigma$ will be normalised to the reference non-stiff case for each power and damping  $(V_0,\alpha_0)$, so that $\sigma>1$ means improved stability by stiffness (e.g. Fig.~\ref{fig::fig4}b). \par
    To fully understand the effects of stiffness, let us first describe the main features of the deposition in the absence of stiffness, as was shown in [\onlinecite{rodr19}]. Let us classify the main changes that deposition suffers into three different stages for a prototypical hysteresis (low $\alpha_0$) solution (a similar division may be suited in cases where no hysteresis exists). The reader might look at the curve labelled $\infty$ in Fig.~\ref{fig::fig4}a. \par
    The first stage (Stage I) occurs for the lower $V_0$ values, and corresponds to the situation before any plasma heating becomes significant. In such a stage, deposition is exponential and thus tends to be larger close to the edge of the island. As damping is weak, significant energy leaks out of the island ($1-V(1)/V(-1)\sim0$). This stage may be related to the lower branch of the hysteresis. \par
    As power increases and the island heats up, preferentially in its centre, the wave starts to self-focus. This moves deposition closer to the centre, and diminishes the power leakage out of the island (Stage II). This corresponds roughly to the transition region between bifurcations (or the large change in Fig.~\ref{fig::fig4}a). \par
    Finally, if the wave power is even larger, then the damping becomes so large that the wave starts to become damped closer to the edge, i.e. shadowing (Stage III). This would correspond to the hotter branch in the hysteresis beyond the bifurcations (and the decreasing part of the $\infty$ curve in Fig.~\ref{fig::fig4}a). In practice, the shadowing can at least be partially compensated by adjusting the aiming of the ray trajectories to force the peak of the nonlinear damping closer to the island centre.   \par
    Because stiffness behaves as a moderator of the non-linear response, it will also moderate the wave self-focusing effect. The effects of stiffness will then be different depending on which of the three stages a given value of $\kappa_c$ affects most, providing a natural grouping. For simplicity of the argument we will assume that the system is stiff enough so that the effects of enhanced transport are significant for all cases beyond the power value associated to a given temperature gradient threshold. A lower stiffness $\widetilde{\chi}_s$ would make the distinction in the grouping of the effects less clear.\par
   
    \par 

    \par
    Consider then a first group of stiffness effects to correspond to those cases for which the cutoff $\kappa_c$ is low enough so as to affect all solutions beyond Stage I. Keeping in mind the relation between $\kappa_c$ and $V_0$, the bound of this group would be that of Stage I, roughly where the non-stiff deposition starts to get closer to the centre of the island (see Fig.~\ref{fig::fig4}a). In this situation, stiffness will freeze the non-linear response of the system before any beneficial condensation effect has taken place. Thus the centring of the deposition or reduction of power leakage are relegated to larger $V_0$ (e.g. compare $|x_\mathrm{pk}|$ for $\kappa_c=1$ in Fig.~\ref{fig::fig4}a to the $\kappa\rightarrow\infty$ case). However, and as Fig.~\ref{fig::fig4}b shows, stability is worsened for the  more relevant lower $V_0$ regime; thus the effects in this group are overall deleterious. 
    \par
    The next group is formed by the set of $\kappa_c$ values that affect Stage II (but not I). Physically, this amounts to stiffness becoming relevant before wave power leakage is reduced completely, but after some significant amount of focusing has occurred. Hence, it is clear that depending on how early stiffness freezes the non-linear behaviour, stability will improve or worsen (see Fig.~\ref{fig::fig4}b). The detrimental contribution is minimal for the larger power situations (as in the previous group), but within this group stability is maximised for larger $\kappa_c$.  \par
    Finally, the larger $\kappa_c$ cases solely affect Stage III correspond to . Naturally, stabilisation is always improved, because at this point stiffness exclusively alleviates the shadowing effect, all other beneficial focusing effects having already taken place. However, if $\kappa_c$ is too large, then only for extremely large powers (relative to $\kappa_c$) will any improvement be tangible. \par

    The balance between the advantageous and detrimental contributions of the non-linear effects that we have gone through may be seen as an optimisation problem. This optimisation has been presented in a simplified form, keeping parameters such as $\alpha_0$ or $\widetilde{\chi}_s$ fixed. The more general multivariate optimisation problem is left for future work, and should be taken into account in designing island stabilisation schemes. \par
    Before concluding, brief contact is made with experimentally relevant values. Typical parameters for stiffness are $\kappa_c\sim5(2)$ and $\chi_s/\chi_0\sim0.01-6$, with $\Upsilon\sim0.1$, as may be seen in Appendix B. These suggest that, generally, turbulent transport will be a relevant process in real devices. We recall here that smaller islands are more susceptible than larger ones, and this will ultimately control when the various effects are relevant. It is left for future work a more complete, fundamental exploration of $\kappa_c$ and $\chi_s/\chi_0$, especially distinguishing the different species in the plasma. In particular, the instability threshold is expected to be less constraining for electrons than for ions. Thus, for slow temperature equilibration times compared to diffusion\cite{suying20}, one could expect to mainly heat electrons, and thus alleviate stiffness constraints. A full description including these species effects, full 3D toroidal geometry and realistic ray tracing is left for future studies. 

\section{Conclusion}
The influence of turbulent enhancement of heat transport on the non-linear condensation effect is explored in this paper. The enhancement occurs when the temperature gradient exceeds a certain threshold, and it is shown to act as an additional island edge localised energy loss mechanism. This way, a very strong stiffness may constrain temperature, giving rise to phenomena similar formally to that in [\onlinecite{rodr19}], even without the need to limit the wave energy. \par
As stiffness is moderated, and using the critical gradient model\cite{Garbet2004}, important changes occur to the non-linear condensation effects. Specifically, bifurcations appear to be modified or destroyed, and generally the response of the system becomes less sensitive to RF power as a result of the stiff moderation. Naturally, islands of smaller size and at larger radius are more likely to exhibit turbulent transport effects. \par 
The location of the RF deposition peak and, more importantly, the stabilisation power are both affected by stiffness. The inhibition of the wave damping can prevent the edge shadowing effect when it occurs, not necessarily compromising the beneficial effects of condensation.\par
Stiffness should therefore be taken into consideration alongside RF condensation in search of optimal island stabilisation schemes. While the model offered here uncovers the key effects, any detailed prediction would require a more complete description, including the species dependence of stiffness, 3D geometry and full ray tracing.

\begin{acknowledgments}
Thanks to Suying Jin for fruitful discussions and help. \par
This work was supported by US DOE DE-AC02-09CH11466 and DE-SC0016072.

\end{acknowledgments}

\appendix

\section{Eigenvalue equation}
Consider equation,
\begin{equation}
    u''=-P_0e^u \label{eqn::app1}
\end{equation}
for constant heat conductivity. Upon integration the equation becomes,
\begin{equation}
    \frac{u'^2}{2}=P_0(\lambda-e^u) \label{eqn::app2}
\end{equation}
where $\lambda$ is some integration constant which will be later interpreted as $\lambda=\exp[u(0)]$. Solving this latter equation exactly, and requiring even symmetry,
\begin{equation}
    u = \ln\lambda-2\ln\left[\cosh\left(\sqrt{\frac{\lambda P_0}{2}}\widetilde{x}\right)\right] \label{eqn::app3}
\end{equation}
The eigenvalue equation simply arises from the boundary condition $u=0$ at $\widetilde{x}=1$,
\begin{equation}
    \lambda=\cosh^2\left(\sqrt{\frac{\lambda P_0}{2}}\right)
\end{equation}
This will hold true everywhere within the island so long as the temperature gradient at the edge of the island does not exceed the threshold gradient. When this is not the case, the eigenvalue equation needs to be modified to accommodate  $\widetilde{\kappa}\rightarrow\infty$ when the threshold $u'_\mathrm{th}$ is overcome.\par
The first thing to solve for is the point $x_\mathrm{E}$ at which the gradient as a solution to Eqn.~(\ref{eqn::app1}) is equal to $u'_\mathrm{th}$. From Eqn.~(\ref{eqn::app2})
\begin{equation}
    1+x_\mathrm{E}=\frac{1}{u'_\mathrm{th}}\ln\left(\lambda-\frac{(u'_\mathrm{th})^2}{2P_0}\right)
\end{equation}
where $\lambda\geq\lambda_*=1+(u'_\mathrm{th})^2/2P_0$ but $\ln[\lambda-(u'_\mathrm{th})^2/2P_0]/u'_\mathrm{th}\leq1$. With this at hand, and noting that the solution (\ref{eqn::app3}) still holds for the central part $(x_\mathrm{E},-x_\mathrm{E})$ of the island and outside $u'=u'_\mathrm{th}$, the eigenvalue equation may be obtained by setting $u(x_\mathrm{E})=u'_\mathrm{th}(1+x_\mathrm{E})$,
\begin{equation}
    \lambda=\cosh^2\left(x_\mathrm{E}\sqrt{\frac{\lambda P_0}{2}}\right)\left(\lambda-\frac{(u'_\mathrm{th})^2}{2P_0}\right)
\end{equation}
Putting all together,
	\begin{align}
	\lambda=&\cosh^2\left(\sqrt{\lambda P_0/2}\right) ~~~ &\left(\lambda\leq\lambda^*\right)\\
	=&\cosh^2\left(x_\mathrm{E}\sqrt{\lambda P_0/2}\right)\left(\lambda-\lambda_*+1\right) ~ &\left(\lambda>\lambda^*\right)
	\end{align}
	
\section{Experimental parameter values}
The critical gradient semi-empirical model for the heat transport is customarily written in the form,\cite{Garbet2004}
\begin{equation}
    \kappa=\kappa_{ 0}\left[1+\frac{\chi_s}{\chi_0}\left(\frac{-R\partial_rT}{T}-\kappa_c\right)H\left(\frac{-R\partial_rT}{T}-\kappa_c\right)\right],
\end{equation}
where $\kappa_{ 0}$ is the non-stiff heat conductivity, $\chi_s/\chi_0$ the strength of the stiffness, $R$ is the major radius, $T$ the temperature, $\kappa_c$ the temperature gradient threshold for stiffness and $H$ the Heaviside function. \par
    To adequately incorporate this temperature gradient dependent conductivity into the governing magnetic island equation, a common set of non-dimensional variables needs to be used. Using the definitions of $\widetilde{x}=2x/W_i$ and $u=w^2\widetilde{T}/T_0$, one may define a new $\widetilde{\kappa}$,
    \begin{equation}
        \widetilde{\kappa}=1+\widetilde{\chi}_s(u'-u'_\mathrm{th})H(u'-u'_\mathrm{th})
    \end{equation}
    where $\widetilde{\chi}_s=\chi_s/\chi_0\Upsilon$ with $\Upsilon=W_iw^2/2R$ and $ u'_\mathrm{th}=\Upsilon\kappa_c$. Typical tokamak values for ECCD schemes will have $\Upsilon\approx5\cdot10~\mathrm{cm}/5~\mathrm{m}\approx10^{-1}$, and will be assumed throughout the text unless otherwise stated.
\begin{table}[h]\vspace{+0.2cm}
\begin{tabular}{c|ccc}
    & $\chi_{s}/\chi_{0}$ & $\kappa_{c}$ \\\hline
    JET &	6(4) &	5.0(1) \\
    AUG & 6(6) & 6(2) \\
    FTU & 0.7 & 8 \\
    TS & 28 & 3 \\
\end{tabular}
\caption{Experimental values of critical gradient electron stiffness model parameters for various shots at Joint European Torus (JET), ASDEX-Upgrade (AUG), Frascati Tokamak Upgrade (FTU) and TORE-SUPRA (TS). Number in brackets represents variation between different shots. See [\onlinecite{Garbet2004}] for more details on the data.} 
\label{table::expParam} 
\end{table} 
\par 
In Table \ref{table::expParam} empirical values for the threshold gradient and stiffness strength from a number of tokamak experiments are collected. It is to be noted that electron stiffness is better documented than ion stiffness, as the temperature of ions is difficult to measure with the needed accuracy. Nevertheless, the quoted values are taken as representative and used in our species-independent picture, which would have to be refined in future work. \par
For guidance on the experimentally relevant values of $V_0$ and $\alpha_0$ in the wave-damping model, we refer the reader to [\onlinecite{rodr19}], where an extensive discussion may be found. Here, we limit ourselves to quoting typical values that may be expected to be $\alpha_0\sim0.1-3$ while $V_0\sim1-10$. 

\section{Stabilisation power $\sigma$}
In this brief Appendix a description of the parameter $\sigma$ suited to describe the stabilisation capability of a given power deposition is given. The formalism used here is heavily based on that used in works such as [\onlinecite{Lazzari}]. \par
Consider a given 1D power deposition solution to understand its stabilisation capability. We would have to compute its contribution to $\Delta'$ in the Modified Rutherford equation. As stability calculations are inherently 2D in the poloidal plane, one needs to interpret the given 1D profile as embedded in the higher dimension. A natural interpretation is to take the deposition as a narrow, O-point centred RF beam, which crosses the elongated island perpendicular to its longer axis. Formally, in the limit of infinitely narrow, let the power be $P(x,\xi)=P(x)\delta(\xi)$, where $x$ is the island width coordinate and $\xi$ corresponds to the helical coordinate (along the longer axis). \par
\begin{figure}[h]
   \hspace*{-0.5cm}
    \includegraphics[width=0.55\textwidth,keepaspectratio]{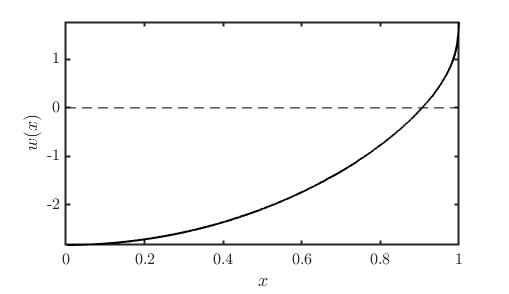}
    \vspace*{-0.7cm}
    \caption{Weight function $w(x)$ for the definition of the stabilisation parameter $\widetilde{\sigma}$. This shows the destabilising nature of areas close to the island edge (roughly a 10\% of the half width).}\label{fig::figApp}
    \end{figure}

Given this set-up, the contribution to $\Delta'$ from inside the island is, ignoring the variation of the current drive efficiency within the island, as well as overall factors,
\begin{equation}
    \Delta'\propto-\int_{-1}^{1}\mathrm{d}\psi\bar{P}(\psi)\int_{-\hat{\xi}}^{\hat{\xi}}\mathrm{d}\xi\frac{\cos\xi}{\sqrt{\psi+\cos\xi}} \label{eqn::delta'}
\end{equation}
where $\psi$ is the flux coordinate, with a particular geometry described by $\psi=2x^2-\cos\xi$ for $m=1$ and $\hat{\xi}=\mathrm{acos}(-\psi)$. Here $x$ is normalised to the island half width. The flux surface average power function $\bar{P}$ is defined as,
\begin{align}
    &\bar{P}(\psi)=\frac{\langle P\rangle}{\langle 1\rangle}\\
   & \langle P\rangle\propto\int_{-\hat{\xi}}^{\hat{\xi}}\mathrm{d}\xi \frac{P(x)+P(-x)}{2\sqrt{\psi+\cos\xi}}\delta(\xi)\propto \frac{P(x)+P(-x)}{2\sqrt{1+\psi}}\\
   & \langle 1\rangle \propto\int_{-\hat{\xi}}^{\hat{\xi}}\frac{\mathrm{d}\xi}{\sqrt{\psi+\cos\xi}}
\end{align}
Therefore, we may re-express the integral in Eqn.~(\ref{eqn::delta'}) to define a stability parameter $\widetilde{\sigma}$
\begin{align}
\begin{split}
    \widetilde{\sigma}\equiv & -\int_{-1}^1 \mathrm{d}\psi \frac{P(x)+P(-x)}{2\sqrt{1+\psi}}\frac{\langle\cos\xi\rangle}{\langle1\rangle} \\
    =& \int_0^1\mathrm{d}x\frac{P(x)+P(-x)}{2}w(x)
    \end{split}
\end{align}
where $w(x=\sqrt{(\psi+1)/2})=-4\langle\cos\xi\rangle/\langle1\rangle\sqrt{2}$ may be interpreted as a sort of weight function (see Fig.~\ref{fig::figApp}). The weight function makes it clear that driving current in the island centre is stabilising, unlike doing it close to the edge. \par
Parameter $\widetilde{\sigma}$ represents stabilisation strength, with a larger negative value representing a more stable power deposition. In general, this metric will be used in relative terms, with a specified normalisation, for which the symbol $\sigma$ is reserved.

\bibliography{stiffPaper}

\end{document}